\documentclass{Interspeech2024}

\usepackage{multirow}
\usepackage{url}
\usepackage{hyperref}
\usepackage{booktabs}
\usepackage{cite}
\usepackage{cleveref}

\newcommand{\astfootnote}[1]{%
  \begingroup
  \renewcommand\thefootnote{\fnsymbol{footnote}}%
  \footnotetext[1]{#1}%
  \endgroup
}
\newcommand{\daggerfootnote}[1]{%
  \begingroup
  \renewcommand\thefootnote{\fnsymbol{footnote}}%
  \footnotetext[2]{#1}%
  \endgroup
}

\interspeechcameraready

\title{Enhancing Zero-shot Audio Classification using Sound Attribute Knowledge from Large Language Models}

\name[affiliation={1}]{Xuenan}{Xu$^{\ast}$}
\name[affiliation={1}]{Pingyue}{Zhang$^{\ast}$}
\name[affiliation={2}]{Ming}{Yan}
\name[affiliation={2}]{Ji}{Zhang}
\name[affiliation={1}]{Mengyue}{Wu$^{\dagger}$}

\address{
  $^1$MoE Key Lab of Artificial Intelligence X-LANCE Lab, Shanghai Jiao Tong University, China\\
  $^2$Institute of Intelligent Computing, Alibaba Group, China
  }
\email{\{wsntxxn, williamzhangsjtu, mengyuewu\}@sjtu.edu.cn, \{ym119608, zj122146\}@alibaba-inc.com}


\begin{document}

\maketitle

\begin{abstract}
Zero-shot audio classification aims to recognize and classify a sound class that the model has never seen during training.
This paper presents a novel approach for zero-shot audio classification using automatically generated sound attribute descriptions.
We propose a list of sound attributes and leverage large language model's domain knowledge to generate detailed attribute descriptions for each class.
In contrast to previous works that primarily relied on class labels or simple descriptions, our method focuses on multi-dimensional innate auditory attributes, capturing different characteristics of sound classes.
Additionally, we incorporate a contrastive learning approach to enhance zero-shot learning from textual labels. 
We validate the effectiveness of our method on VGGSound and AudioSet\footnote{The code is available at \url{https://www.github.com/wsntxxn/AttrEnhZsAc}.}.
Our results demonstrate a substantial improvement in zero-shot classification accuracy.
Ablation results show robust performance enhancement, regardless of the model architecture.
\end{abstract}

\section{Introduction}
\astfootnote{Equal contribution.}
\daggerfootnote{Corresponding author.}
Supervised learning has shown promise in many fields however it is limited to pre-defined classes included during training. 
In audio classification, current datasets often only contain commonly-heard daily sounds~\cite{gemmeke2017audio} while rare sounds from specific domains are seldom included due to requirement of labor-intensive annotations. 
Hereby, such models can only predict the probability of pre-defined classes but the probability of unseen classes cannot be given, impeding its real-world applications.

To overcome the reliance on exhausting annotation of training data, zero-shot classification has drawn much attention recently~\cite{primus2022improved,sims2023enhanced}.
Most works learn the correspondence between audio features and semantic features of classes~\cite{gudur2021zero,choi2019zero}.
Xie \textit{et al.} used a bilinear model to calculate the audio-class similarity~\cite{xie2021zeroicassp,xie2021zerotaslp}.
The textual label is the class and a definition from wikipedia, \emph{e.g.}, ``Sounds associated with the species of medium-to-large birds, Corvus'' for the class \texttt{Crow}.
Sims \textit{et al.} added synonyms, semantic broadening and onomatopoeia as the auxiliary information~\cite{sims2023enhanced}.
However, labels and auxiliary information are often insufficient to discriminate sounds, in particular similar ones.
Since the definition is about the sounding object instead of auditory attributes, they cannot reflect the difference between close-related classes, e.g., \textit{train} and \textit{underground}.
Therefore, these approaches have not extensively exploited perceptual auditory knowledge to describe sound classes.

Auditory attributes, more representative of sounds' inherent characteristics, can be a leading flag for sound descriptions.
Some studies explored using sound attributes as either inputs or training objectives to facilitate learning: 
Lin \textit{et al.} used attributes like pitch and material to characterize collision sounds~\cite{lin2023zero} while
Choi \textit{et al.} took the occurrence of instruments as attributes to classify music.


Recently, audio foundation models from contrastive language-audio pre-training (CLAP)~\cite{elizalde2023clap,wu2023large} have drastically improved the generalizability of audio models by using natural language supervision.
However, the success in effective zero-shot classification may be due to indirect exposure to testing sounds during training, rather than true zero-shot learning.
For instance, CLAP's ability to recognize crow cawing could be attributed to its training on paired data of crow cawing sounds and the text ``A crow is cawing'', though the sound is not explicitly labelled as a pre-defined class.
Thus, CLAP's capability of classifying novel sound categories without prior exposure may be questioned.
Moreover, CLAP predominantly focuses on scaling-up audio-text data while the text data are mostly limited to event labels.
These textual descriptions may not adequately capture the distinct characteristics inherent in auditory signals, such as pitch, tone, and timbre.
Therefore, we adopt the zero-shot setting that test audio is unseen during training and focus on improving textual descriptions in this work. 

\begin{table*}[!ht]
    \centering
    \begin{tabular}{c|l}
        \hline\hline
        \textbf{Attributes} & Remarks: Take ``\textit{Smoke detector, smoke alarm}'' as an example \\
        \hline
        \textbf{Sound class} & The name of the sound class is one of the attributes: \textit{Smoke detector, smoke alarm}. \\
        \hline
        \textbf{Sound frequency (pitch)} & The frequency of the sound: \textit{frequency is sharp, piercing, and high}.\\
        \hline
        \multirow{2}*{\textbf{Timbre}} & The timbre of the sound, that is, the unique quality or character of a sound: \\
        & \textit{timbre is shrill and alarming}. \\
        \hline
        \textbf{Onomatopoeia} & Words to imitate or resemble the sound: \textit{sounds like beep, beep-beep, and loud screech}\\
        \hline
        \textbf{Simile} & Comparison to another common sounds: \textit{like a persistent and urgent warning signal}.\\
        \hline
        \textbf{Emotion} & The emotions evoked by the sound: \textit{evoking alertness and urgency}. \\
        \hline
        \multirow{2}{*}{\textbf{Definition}} & Only available for AudioSet, a wikipedia description of the sound: \\
        & \textit{Sounds emitted by a device that senses smoke, typically to warn of a fire}. \\
        \hline
    \end{tabular}
    \caption{Seven attributes defined to describe sound, using smoke detector as the example. To differentiate the ``description'' provided by AudioSet from our proposed attribute-oriented description, we use ``definition'' to denote the former.
    }
    \vspace{-4mm}
    \label{tab:attribute}
\end{table*}

In this paper, we integrate the advantages of semantic supervision and attribute supervision to improve zero-shot audio classification by learning the correspondence between audio and its description that focuses on sound attributes.
Compared with previous works using sound attributes, we define a set of attributes applicable to general sound categories.
Then we use ChatGPT to automatically generate attribute-focused descriptions for all categories.
The capabilities of large language models (LLM) are exploited to provide accurate attribute descriptions without human annotation.
Moreover, inspired by the success of self-supervised learning~\cite{khosla2020supervised}, we adapt a more advanced contrastive learning paradigm to zero-shot classification, thereby achieving improved alignment between audio and textual labels.
Note that our approach is knowledge-driven, hence can be integrated to foundation models like CLAP by adding the sound attributes and improving textual labels during training.
Experimental results demonstrate that our approach achieves significant improvement in zero-shot classification accuracy, regardless of the backbone model architecture. 
In summary, our contributions are three-fold: 
\begin{itemize}
\setlength{\itemsep}{0pt}
    \item We define an inherent auditory attribute set for general sounds and incorporate ChatGPT~\cite{openai2021chatgpt} to automatically generate attribute-oriented descriptions, which can assist any audio model architecture training. 
    \item We use an improved contrastive learning paradigm to learn better audio-text correspondence.
    \item Results on two datasets demonstrate that by leveraging the audio-related domain-specific knowledge in LLM mined from massive text data, the zero-shot classification performance is significantly improved.
\end{itemize} 

\section{Description of Sound Attributes}
\label{sec:attribute_chatgpt}

To describe the characteristics of a sound class, it is necessary to define a set of sound attributes.
Although \cite{lin2022binary} defined several binary attributes such as duration and material, these attributes only cover a small portion of sound features and are somewhat only suitable for the involved collision sound dataset.
Therefore, we try to determine general sound attributes.
We ask ChatGPT which attributes can be used to describe a sound category.
In addition to the summarized attributes, we manually add three attributes:
1) onomatopoeia: inspired by \cite{sims2023enhanced}, we include onomatopoeia to supplement the description of sounds that are difficult to describe using attributes such as pitch and intensity;
2) timbre: a property that distinguishes different sounds, especially those with similar pitches;
3) simile: similar sounds.
In zero-shot classification, the model can be extensively trained on seen categories.
Therefore, when calculating the probability of an unseen class, information about what it is similar to among the seen classes is valuable.

Based on these attributes, we first ask ChatGPT\footnote{We use \textit{gpt-3.5-turbo}.} to generate descriptions for several sound classes.
Subsequently, we take them as few-shot examples to guide ChatGPT to generate descriptions for the remaining classes.
Such an in-context learning approach makes full use of the extensive acoustic knowledge learned by the large language model.
Finally, descriptions in terms of 6 attributes are generated for each sound class, as shown in \Cref{tab:attribute}.
For AudioSet, we also use its provided wikipedia description as an additional attribute.

\section{Zero-Shot Classification by Audio-Text Contrastive Learning}
\label{sec:method}

\begin{figure*}[htpb]
    \centering
    \includegraphics[width=0.9\linewidth]{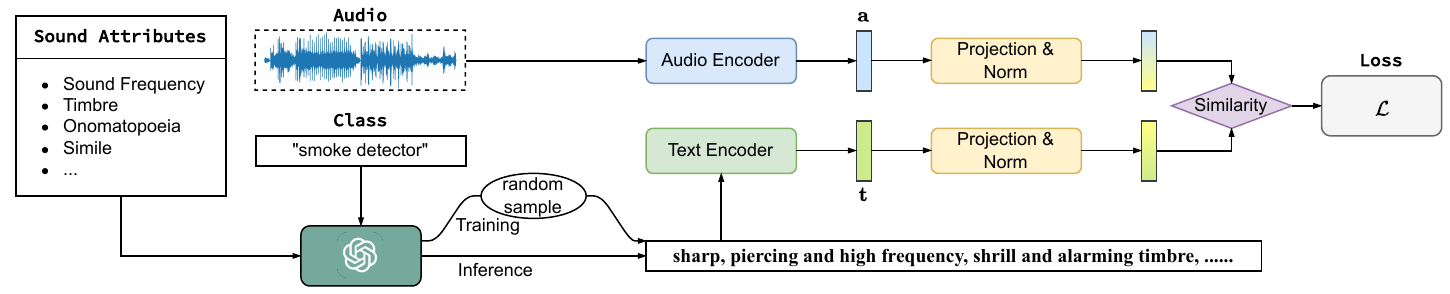}
    \caption{The contrastive learning framework for zero-shot audio classification. The text is the description of sound attributes for the class generated by ChatGPT. During training, attributes are randomly selected to form the description while all attributes are used during inference.}
    \label{fig:approach}
\end{figure*}

\subsection{Contrastive Learning Framework}
Our zero-shot classification approach is similar to previous studies~\cite{xie2021zeroicassp,sims2023enhanced}, with a few modifications.
The overall framework is shown in \Cref{fig:approach}.
For a pair of an audio clip $\mathcal{A}$ and a class label $c$, sound attributes $\mathcal{T}_1, \mathcal{T}_2, \cdots, \mathcal{T}_n$ of $c$ are annotated by ChatGPT.
An audio encoder transforms the input audio $\mathcal{A}$ into an embedding vector $\mathbf{a}$:
\begin{equation}
    \mathbf{a} = \mathrm{Enc}_A (\mathcal{A})
\end{equation}
During training, the descriptive text for an audio sample is formed by sampling from $\mathcal{T}_1, \mathcal{T}_2, \cdots, \mathcal{T}_n$ and concatenation.
Different sampling strategies are investigated and the comparison will be shown later in \Cref{subsec:ablation}.
The description is encoded by a text encoder accordingly:
\begin{equation}
    \mathbf{t} = \mathrm{Enc}_T (\mathrm{Concat}(\mathcal{T}_1, \mathcal{T}_2, \cdots, \mathcal{T}_s))
\end{equation}
where $s$ is the number of sampled attributes.

The model is trained to minimize the distance between $\mathbf{a}$ and its corresponding class's $\mathbf{t}$, while repelling $\mathbf{a}$ and $\mathbf{t}$ of other classes, using a similarity function $\mathrm{sim}(\cdot, \cdot)$.
During inference, all attributes of the class $c$ are utilized to calculate the similarity:
\begin{equation}
    \mathbf{t}^c = \mathrm{Enc}_T\left(\mathrm{Concat}(\mathcal{T}_1, \mathcal{T}_2, \cdots, \mathcal{T}_n)\right)
\end{equation}
\begin{equation}
    y = \underset{c \in C}{\mathrm{argmax}} \ \mathrm{sim}(\mathbf{a}, \mathbf{t}^c)
\end{equation}
where $C$ is the set of zero-shot classes for testing which does not intersect with the training class set.
Compared with the framework in \cite{xie2021zerotaslp,lin2023zero}, our approach make two modifications:

\paragraph*{Similarity Function}
The baseline in \cite{xie2021zerotaslp} calculates audio-class similarity by a bilinear model:
\begin{equation}
    s = (W'\mathbf{a})'\mathbf{t}
\end{equation}
We replace it with the cosine similarity between embeddings after fully-connected (FC) projection layers:
\begin{align}
    \begin{split}
    \mathbf{a}_P = \mathrm{FC}_A(&\mathbf{a}) \quad \mathbf{t}_P = \mathrm{FC}_T(\mathbf{t})\\
    s = &\frac{\mathbf{a}_P \cdot \mathbf{t}_P}{\lVert \mathbf{a}_P \rVert \cdot \lVert \mathbf{t}_P \rVert}
    \end{split}
\end{align}

\paragraph*{Loss Function}
In \cite{xie2021zerotaslp} the model is trained by a weighted max margin ranking loss~\cite{weston2011wsabie}.
In light of the success of InfoNCE loss~\cite{oord2018representation} in supervised contrastive learning (SupCon)~\cite{khosla2020supervised}, we adopt it in our improved learning framework.
For a batch of $B$ samples, the loss is calculated as:
\begin{equation}
     \mathcal{L} = \sum_{i} \frac{-1}{N_{l_i}}\sum_{j, l_j = l_i} \log \frac{\exp\left(\mathrm{sim}(\mathbf{a}_i, \mathbf{t}_j) / \tau\right)}{\underset{k}{\sum} \exp{\left(\mathrm{sim}(\mathbf{a}_i, \mathbf{t}_k) / \tau\right)}}
\end{equation}
where $i, j, k \in \{1...B\}$, $l_i$ is the label of the audio sample $i$ and $\tau$ is the temperature.
$N_{l_i}$ is the number of samples in the batch that share the same label with $i$.
Descriptions from these samples are treated as positive ones while other samples' descriptions are negative.
The InfoNCE loss trains the model to discriminate positive descriptions from negative ones.

To sum up, our primary approach involves two methods: \textbf{Baseline} and \textbf{SupCon}.
In \textbf{Baseline}, we utilize bilinear similarity and employ the weighted max-margin ranking loss. 
This method is the same as \cite{xie2021zerotaslp}, using all attributes as the textual label.
\textbf{SupCon} combines cosine similarity with the InfoNCE loss as done in \cite{khosla2020supervised}.
In this method, we randomly select attributes during training.

\subsection{Model Architecture}

To encode informative attribute descriptions into discriminative embeddings, we employ SentenceBERT~\cite{reimers2019sentence} as the text encoder.
It is capable of encoding sentences into discriminative embeddings, thus well-suited for distinguishing descriptions of different sound classes.
We explore two variants of SentenceBERT: a standard \texttt{mpnet-base} with 12 layers and a hidden size of 768, and a smaller \texttt{MiniLM} with 6 layers and a hidden size of 384.
Different text encoders are evaluated to investigate the correlation between the performance of our approach and the encoding ability of the text encoder. 
We validate the effectiveness of our method on 2 popular audio encoder backbones: a convolutional PANNs \texttt{CNN14}~\cite{kong2020panns} and a Transformer-based \texttt{AST} tiny224~\cite{gong21b_interspeech}.
Since the size of audio classification datasets is insufficient to fine-tune the pre-trained text encoder, we keep the text encoder frozen while the audio encoder is trained from scratch.



\section{Experimental Setup}
\label{sec:exp_setup}

\paragraph*{Dataset}
Although previous works on zero-shot audio classification mostly used ESC50~\cite{piczak2015esc}, we find results on ESC50 unstable due to limited data size.
Therefore, we use two large datasets, VGGSound~\cite{chen2020vggsound} and AudioSet~\cite{gemmeke2017audio}. 
VGGSound contains over 200k clips 
covering 309 classes. 
For AudioSet, we select only single-label audio samples from unbalanced set and exclude 21 abstract classes like ``Animal'' and ``Music''.
We also exclude classes with less than 100 samples.
Following \cite{xie2021zerotaslp}, we randomly select 1,500 samples for classes with more than 1,500 samples for a balanced distribution.
Finally, 115k samples and 285 classes are selected for AudioSet.

\paragraph*{Fold Split}
To ensure fully zero-shot setting, we split a dataset into 5 disjoint folds according to classes. 
The model is then trained on 4 of them and evaluated on the remaining one. 
For a balanced data distribution among folds, we sort classes in descending order based on their sample numbers.
Starting with 5 empty folds, we select the next 5 unallocated classes in the sorted order, shuffle them, and allocate them to the 5 folds respectively.
The splitting process yields approximately 35k samples in each fold for VGGSound and 23k samples for AudioSet.


\begin{table*}[ht]
    \centering
    \caption{Zero-shot classification performance using CNN14 backbone. ``Definition'' here refers to wikipedia description only available for AudioSet.
    All attributes are used during training for both approaches. 
    }
    \begin{tabular}{c|c|ccccc|ccccc}
    \toprule
    \multicolumn{1}{c}{} & \multicolumn{1}{c}{} & \multicolumn{5}{|c|}{VGGSound} & \multicolumn{5}{c}{AudioSet} \\
    \midrule
    Approach & Text & Fold 1 & Fold 2 & Fold 3 & Fold 4 & Fold 5 & Fold 1 & Fold 2 & Fold 3 & Fold 4 & Fold 5 \\
    \midrule
    \multirow{2}{*}{Baseline~\cite{xie2021zeroicassp}} & Class (\& Definition) & 20.5 & 19.9 & 23.7 & 20.8 & 21.4 & 23.7 & 17.7 & 24.2 & 26.1 & 23.6 \\
     & Attributes & 24.1 & 22.6 & 27.1 & \bf{25.6} & 22.5 & 25.2 & 23.5 & 26.6 & 28.9 & 26.8 \\
    \midrule
    \multirow{2}{*}{SupCon} & Class (\& Definition) & 23.2 & 21.7 & 23.5 & 21.6 & 24.2 & 25.8 & 21.6 & 27.0 & 27.9 & 27.8 \\
     & Attributes & \bf{27.0} & \bf{25.3} & \bf{27.3} & 25.3 & \bf{26.6} & \bf{26.1} & \bf{26.0} & \bf{29.0} & \bf{31.8} & \bf{32.5} \\
    \bottomrule
    \end{tabular}
    \label{tab:results}
\end{table*}



\paragraph*{Hyper-parameters}
Following \cite{xie2021zerotaslp}, we apply a two-stage training paradigm: 1) training the audio encoder from scratch; 2) performing \textbf{Baseline} or \textbf{SupCon} audio-text alignment.
The model is trained with a batch size of 64 using SGD optimizer with an initial learning rate of 0.1 and the cosine learning rate scheduler in the first stage. Note that in the first stage, the model is also trained on 4 folds and has no access to the leave out fold.
In the second stage, we use Adam optimizer with a learning rate of $1\times10^{-4}$ and a batch size of 256.

\section{Results and Analysis}
\label{sec:results_analysis}

\subsection{Comparison with Baselines}
Results on VGGSound and AudioSet are presented in \Cref{tab:results}.
To achieve fair comparison between \textbf{Baseline} and \textbf{SupCon}, we report results of \textbf{SupCon} with all attributes used during training, which is consistent with \textbf{Baseline}.
We provide accuracy averaged over 5 random seeds.
The standard deviation is omitted since it is minor.
Results of t-tests demonstrate that in all cases where the proposed method outperforms the baseline, the improvement is statistically significant with p-values less than 0.05.
Note that ``definition'' in \Cref{tab:attribute} is unavailable for VGGSound.
Therefore, in terms of the textual label, we solely use class as the baseline for VGGSound, while class and definition for AudioSet.
Results demonstrate the efficacy of using descriptions focusing on sound attributes we define for supervision.
Whatever the textual labels and dataset used, \textbf{SupCon} approach achieves consistent improvement over \textbf{Baseline}.
The comparison validates the effectiveness of our improved contrastive learning paradigm.

\subsection{Ablation Study}
\label{subsec:ablation}


\paragraph*{Attribute Sampling Strategy}

During training, attributes can be sampled with different strategies to form the description text.
We explore three strategies: \textit{Deterministic}, \textit{Random}, and \textit{With-class}.
\begin{itemize}
    \item \textit{Deterministic}: all attributes are selected.
    \item \textit{Random}: attributes are randomly selected.
    \item \textit{With-class}: attributes are also randomly selected but we ensure the class name is always sampled.
\end{itemize}
By default we use the \textit{Deterministic} strategy.
Results in \Cref{tab:ablation_random} showcase the superiority of the \textit{With-class} strategy.
This indicates that class label is a vital and clean attribute hence it should be always included.
Other attributes are generated automatically with inevitable noise. 
Therefore, randomness in attributes other than class should be incorporated into the training process to enhance the model's robustness against noisy attributes.



\begin{table}[ht]
    \centering
    \caption{VGGSound zero-shot classification performance averaged on 5 folds using different attribute sampling strategies.}
    \begin{tabular}{c|ccc}
    \toprule
    Strategy & Deterministic & Random & With-class \\
    \midrule
    Accuracy  & 26.3 & 25.3 & \bf{26.6}  \\
    \bottomrule
    \end{tabular}
    \label{tab:ablation_random}
\end{table}

\paragraph*{Model Architecture}
To investigate the performance of our method under different model architectures, we change the audio / text encoder backbone. Here we use the \textit{With-class} strategy during training.
Results are presented in \Cref{tab:ablation_arch}.
Although AST performs worse than CNN14, the supervision of attribute descriptions does bring improvement over the textual label of class, indicating that our method is robust to audio backbones.
In addition, by replacing the MiniLM text encoder with mpnet which performs better in text matching, further performance boost is achieved.
Our approach is capable of leveraging advanced text encoders to achieve better zero-shot classification performance.

\begin{table}[ht]
    \centering
    \footnotesize
    \caption{VGGSound zero-shot classification performance averaged on 5 folds using different audio / text backbones.}
    \begin{tabular}{c|ccc}
    \toprule
    Text & CNN14-MiniLM & AST-MiniLM & CNN14-mpnet \\
    \midrule
    Class  & 22.8 & 17.9 & 24.1 \\
    Attributes & \bf{26.6} & \bf{21.6} & \bf{28.1} \\
    \bottomrule
    \end{tabular}
    \label{tab:ablation_arch}
\end{table}


\subsection{Analysis on Attributes}


\begin{figure}[ht]
    \centering
    \includegraphics[width=0.95\linewidth]{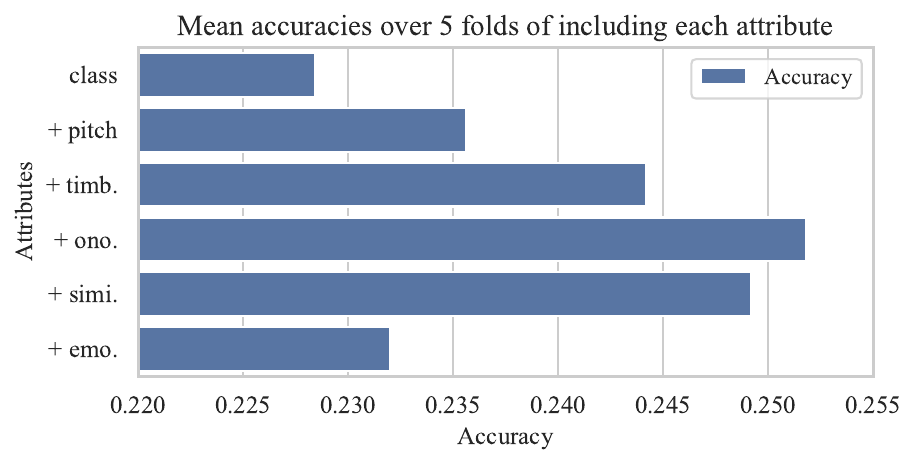}
    \caption{VGGSound zero-shot classification performance with each attribute included during training.}
    \label{fig:ablation_with_attr}
\end{figure}

To analyze the improvement brought by attribute descriptions, we include each attribute during training to explore its influence.
The result in shown in \Cref{fig:ablation_with_attr}. 
Whatever attribute is included, the performance is consistently enhanced.
Compared with low-level auditory attributes like pitch and timbre, high-level semantic attributes such as onomatopoeia and simile bring more significant improvement.
This indicates that high-level semantic attributes generated by LLMs are more accurate and representative for audio classification.

\begin{figure}[htpb]
    \centering
    \includegraphics[width=0.95\linewidth]{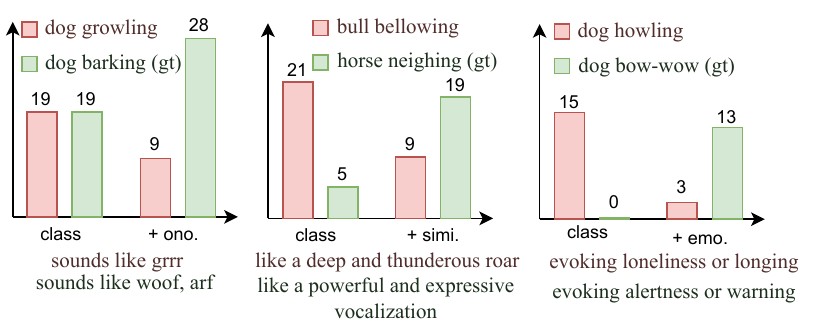}
    \caption{Performance enhancement brought by adding attributes.
    Green and red denote ground truth and misclassified classes.
    The corresponding attribute description is also presented.
    Ono.= onomatopoeia, simi.=simile, emo.=emotion.}
    \label{fig:ablation_attr}
\end{figure}

Then, some cases are selected to show the effectiveness of adding specific attributes, shown in \Cref{fig:ablation_attr}.
Numbers on the green and red bars show numbers of correctly and mistakenly classified samples, respectively.
These classes are easily misclassified for they share similar acoustic characteristics or their labels are semantically close.
Additional attributes describe unique acoustic characteristics to distinguish similar classes.
For example, onomatopoeia is utilized to imitate the sound of dog growling and barking to highlight the difference.

\section{Conclusion}
\label{sec:conclusion}
In this paper, we propose a zero-shot audio classification approach using sound attribute descriptions.
The approach combines the advantages of semantic supervision and attribute supervision by making use of rich knowledge from ChatGPT.
We define a set of sound attributes and use ChatGPT to automatically describe a sound category in terms of each attribute.
Moreover, we propose an improved contrastive learning paradigm to augment the capability of our model to learn effectively from textual labels.
The zero-shot accuracy on VGGSound and AudioSet is significantly boosted, indicating the effectiveness of our approach.
The ablation study on backbones shows consistent improvement over the baseline, validating that our method is agnostic to model architectures.
The limitation is that the attributes described by ChatGPT are occasionally misaligned with their respective attribute names. Additionally, the exploration of utilizing different LLMs for generation is left for future investigation.

\section{Acknowledgements}
This work was supported by National Natural Science Foundation of China (Grant No.92048205), the Key Research and Development Program of Jiangsu Province~(No.BE2022059), Guangxi major science and technology project~(No. AA23062062) and Alibaba Innovative Research.

\bibliographystyle{IEEEtran}
\bibliography{refs}

\end{document}